\documentclass[]{emulateapj}
\usepackage{natbib}

\shorttitle{Comparing SSP models with SDSS cluster colors}
\shortauthors{Peacock et al.}

\begin{document}

\title{Testing stellar population synthesis models with Sloan Digital Sky Survey colors of M31's globular clusters}

\author{Mark B. Peacock and Stephen E. Zepf}
\affil{Department of Physics and Astronomy, Michigan State University, East Lansing, MI 48824, USA}
\email{MBP: mpeacock@msu.edu}

\author{Thomas J. Maccarone}
\affil{School of Physics and Astronomy, University of Southampton, Southampton, SO17 1BJ, UK}

\and 

\author{Arunav Kundu} 
\affil{Eureka Scientific, Inc., 2452 Delmer Street, Suite 100 Oakland, CA 94602, USA}

\begin{abstract}
\label{sec:abstract}

Accurate stellar population synthesis models are vital in understanding the properties and formation histories of galaxies. In order to calibrate and test the reliability of these models, they are often compared with observations of star clusters. However, relatively little work has compared these models in the \textit{ugriz} filters, despite the recent widespread use of this filter set. In this paper, we compare the integrated colors of globular clusters in the Sloan Digital Sky Survey (SDSS) with those predicted from commonly used simple stellar population (SSP) models. The colors are based on SDSS observations of M31's clusters and provide the largest population of star clusters with accurate photometry available from the survey. As such, it is a unique sample with which to compare SSP models with SDSS observations. From this work, we identify a significant offset between the SSP models and the clusters' \textit{g-r} colors, with the models predicting colors which are too red by \textit{g-r}$\sim$0.1. This finding is consistent with previous observations of luminous red galaxies in the SDSS, which show a similar discrepancy. The identification of this offset in globular clusters suggests that it is very unlikely to be due to a minority population of young stars. The recently updated SSP model of Maraston $\&$ Stromback better represents the observed \textit{g-r} colors. This model is based on the empirical MILES stellar library, rather than theoretical libraries, suggesting an explanation for the \textit{g-r} discrepancy.

\end{abstract}


\section{Introduction}
\label{sec:intro}

Stellar population synthesis (SPS) models are an important astrophysical tool for interpreting photometric and spectroscopic observations of galaxies. Modern surveys can detect vast numbers of galaxies across a range of redshifts and can provide important observational constraints on star formation histories and galaxy evolution. However, it is generally not possible to resolve the stellar populations of these galaxies. Instead, their properties are often estimated by comparing their integrated emission with that predicted from composite stellar population models. Such models combine our knowledge of stellar evolution with stellar spectral libraries to predict the integrated emission of an unresolved stellar population. In their simplest form, simple stellar population (SSP) models, SPS models predict the integrated emission from a stellar population with a single metallicity and star formation epoch. The ability of SPS models to accurately represent real stellar populations has a direct influence on galactic studies.

For many years, star clusters have provided some of the best locations to calibrate parameters in SPS models (not all of which can be derived theoretically) and to test their ability to reproduce the observed properties. This is because the relatively simple stellar populations of these clusters can be directly compared with such models. Photometric and spectroscopic observations of the Milky Way's (old) globular clusters, (young) open clusters and the Large/Small Magellanic Cloud's clusters have previously been used to calibrate and test such models \citep[e.g.][]{Renzini88,VandenBerg03,Maraston05,An08,Lyubenova10,Riffel11}. These studies, which often consider observations obtained in the Johnson-Cousins UBVR$_{c}$I$_{c}$ filters, have shown reasonable agreement between models and observations. However, several discrepancies, most notably in the B-V \citep[e.g.][]{Worthey94,Maraston05} and near-infrared \citep[e.g.][]{Cohen07,Conroy10} colors, have been demonstrated.

Despite the increasing popularity and use of the SDSS's \textit{ugriz} filter set \citep{Fukugita96}, relatively little work exists which compares SSP models to star clusters in these filters. This is due primarily to a relative lack of cluster photometry through these, compared with the Johnson-Cousins, filters. However, there have been some suggestions of offsets between SPS models and SDSS observations. In particular, studies of luminous red galaxies (LRGs, which are expected to have a predominantly old, metal rich, stellar population) have shown significant offsets from the model predictions, particularly around the \textit{gri}-bands \citep[e.g.][]{Eisenstein01,Wake06}. Several explanations have been proposed to rectify this offset. \citet{Maraston09} proposed that slight discrepancies between theoretical and empirical stellar libraries may explain the observed offset. They found that models which include a minority population of metal poor stars (3$\%$ of the total) and that are based on empirical, rather than theoretical, libraries better represent the observed colors at all redshifts considered. However, \citet{Conroy10} were unable to explain the LRG colors using these different stellar libraries. Instead, it has also been suggested that a small population of younger or relatively metal poor stars in these galaxies could explain the SDSS observations. The addition of other stellar populations such as blue/extreme horizontal branch (HB) stars and blue straggler stars (typically excluded in such models) could also help to explain the offsets. Some discrepancies have also been suggested between SSP models and the (resolved) SDSS color magnitude diagrams of 17 of the Milky Way's globular clusters \citep{An08}. By comparing their fiducial sequences with the theoretical isochrones of \citet{Girardi04}, \citet{An08} demonstrated that the models could not be simultaneously fit to both the main sequence and red giant branch accurately. 

In this paper, we compare SSP models with the integrated colors of globular clusters in M31. This cluster system is the only large sample of clusters for which accurate integrated photometry is available from the SDSS survey. It therefore provides a unique location in which to compare SDSS photometry of star clusters with SSP models. In section 2 we describe the (previously published) data available for M31's clusters. The SPS models considered in this paper are introduced in section 3. Section 4 compares these models with the cluster data and section 5 considers possible explanations for observed offsets.

\section{M31 globular cluster data}
\label{sec:m31gc}

M31 hosts the largest cluster population in our local group. The galaxy's proximity to us ($\sim$780 kpc) means its clusters' stellar populations are unresolved in typical ground based observations. However, the integrated emission of these clusters can be accurately measured with relatively short exposures (compared with cluster systems around more distant galaxies). In this paper, we consider all globular clusters listed as `class 1' in the catalog of \citet{Peacock10}\footnote{available from the VizieR archive}. These clusters have non-stellar surface brightness profiles and are confirmed to be at the distance of M31 based on spectroscopy or high resolution HST observations. This dataset is therefore unlikely to contain significant contamination from non-cluster sources. This sample also excludes the recently identified population of young clusters in M31. Young clusters are classified as those clusters having colors bluer than the old Milky Way globular clusters \citep{Fusi_Pecci05,Peacock10} or based on spectroscopy of the clusters \citep{Caldwell09}. The catalog includes 416 class 1 clusters, located across the galaxy.

The SDSS covered a large region of sky in the direction of M31 as part of a supplemental run to the main survey. These observations have previously been used to provide \textit{u} and \textit{griz}-band photometry for 73$\%$ and 92$\%$ of M31's known globular clusters, respectively (fewer clusters are detected in the \textit{u}-band due to the increased extinction and lower sensitivity of the survey at these wavelengths). The photometry was performed using \textsc{Sextractor}. The colors were obtained through 4$\arcsec$ apertures and total magnitudes measured using an aperture in the range 2.8-10.6$\arcsec$, depending on the size of the cluster. The photometry was calibrated using the standard SDSS pipeline calibration. As such, the colors are on the AB magnitude system. The SDSS calibration is thought to be in good agreement with this system for the \textit{griz}-bands, although a slight correction has been proposed for the \textit{u}-band \citep[such that $u_{\rm{AB}}$=$u_{\rm{SDSS}}$-0.04;][]{Bohlin01}. For full details of these data we refer the reader to \citet{Peacock10}.

Spectroscopic metallicities for 200 of these clusters are available from the collated catalog of \citet{Fan08}. These measurements are taken from the studies of \citet{Huchra91}, \citet{Barmby00} and \citet{Perrett02}. Where clusters are present in more than one of these studies, preference was given to the larger and more recent catalog of \citet{Perrett02}. \citet{Fan08} also provide estimates for the reddening of these clusters. The extinction in each band was calculated for each cluster using the \citet{Fan08} reddening values and the extinction curves of \citet{Cardelli89}, as presented in table 6 of \citet{Schlegel98}. The reddening of individual clusters in M31 is variable across the galaxy. Therefore, we only consider those clusters for which reddening estimates are available. This results in a sample to 200 of M31's globular clusters which have photometry, spectroscopic metallicities and reddening estimates. To ensure the cluster colors used are relatively robust to variations between the true and assumed extinction curves along their line of sight, we only consider relatively low extinction clusters, with E(B-V)$<$0.2. Our final sample is comprised of 140 clusters.

\section{Simple stellar population models}
\label{sec:SSPs}

In this paper, we consider three of the commonly used SPS models, as well as some recent updates to these models. We summarize these below, but for details of the models, we refer the reader to the cited papers. 

Firstly, we consider the models of \citet[][hereafter BC03]{Bruzual03}\footnote{taken from: http://www2.iap.fr/users/charlot/bc2003/galaxev $\_$download.html}. BC03 provide models based on three stellar evolutionary prescriptions. Of these options we take the models based on the `Padova 1994' tracks. This model combines these tracks with either the STELIB or BaSeL 3.1 stellar libraries. The colors of these models (as a function of age and metallicity) in the SDSS filters are taken from the tables provided by BC03. Models based on two different initial mass functions (IMFs) are available. Only a small difference is observed between the different IMFs, however, the model with a Chabrier IMF provides a slightly better fit to our data than the Salpeter IMF. We therefore select the model with this IMF. 

We also consider the models of \citet[][hereafter M05]{Maraston05}\footnote{taken from: http://www.icg.port.ac.uk/$\sim$maraston/Claudia's $\_$Stellar$\_$Population$\_$Model.html}. These models are based on the theoretical BaSeL stellar atmospheres \citep{Kurucz79,Lejeune98} combined with models for M~giants from \citet{Bessell89} and empirical colors for the thermally-pulsating asymptotic giant branch (TP-AGB) stars. M05 also contains some treatment of the HB star component, providing models for stellar populations with either red or blue HB morphologies. The \textit{ugriz} photometry for these models were taken from the tables provided. As demonstrated later, the model which contains a blue-HB morphology and Kroupa IMF provides the best fit to our data and is the primary model used in our analysis. 

In addition to M05, we also consider the updated models of Maraston $\&$ Stromback (MNRAS submitted,2011). These models are based on different stellar libraries to those presented in M05. Unfortunately, these libraries cover a more limited parameter space than the Kurucz library. The model based on the STELIB library of \citet{LeBorgne03} only includes [Fe/H]$\geq$0.01, and is only applicable to metal rich globular clusters. The MILES libraries of \citet{Sanchez-Blazquez06} and the ELODIE library of \citet{Prugniel07} provide better metallicity coverage but only cover a limited wavelength range, meaning that they are primarily applicable to the SDSS $g$ and $r$-bands. We consider only the updated models based on the MILES and ELODIE libraries (hereafter referred to as MS11-MILES and MS11-ELODIE). The photometry for these models, as a function of age and metallicity, was tabulated by convolving the SDSS filter responses with the model SEDs. 

Finally, we consider models based on the latest Padova tracks and obtained from the web interface CMD 2.3, at http://stev.oapd.inaf.it/cmd. We extracted two sets of models in the ``SDSS \textit{ugriz}" filters. The first model is that of \citet{Marigo08} based on the tracks of \citet{Girardi00,Girardi02}. This includes the bolometric corrections for carbon stars of \citet{Loidl01} and was calculated with no circumstellar dust. We refer to this as the Padova model. The second model is based on the same tracks, but with a correction applied to the TP-AGB tracks based on the new empirical calibration of \citet[][their case B correction]{Girardi10}. We refer to this as the corrected Padova model.

\section{SSP and globular cluster colors}
\label{sec:comparison}

\begin{figure*}
 \centering
 \includegraphics[height=170mm,angle=270]{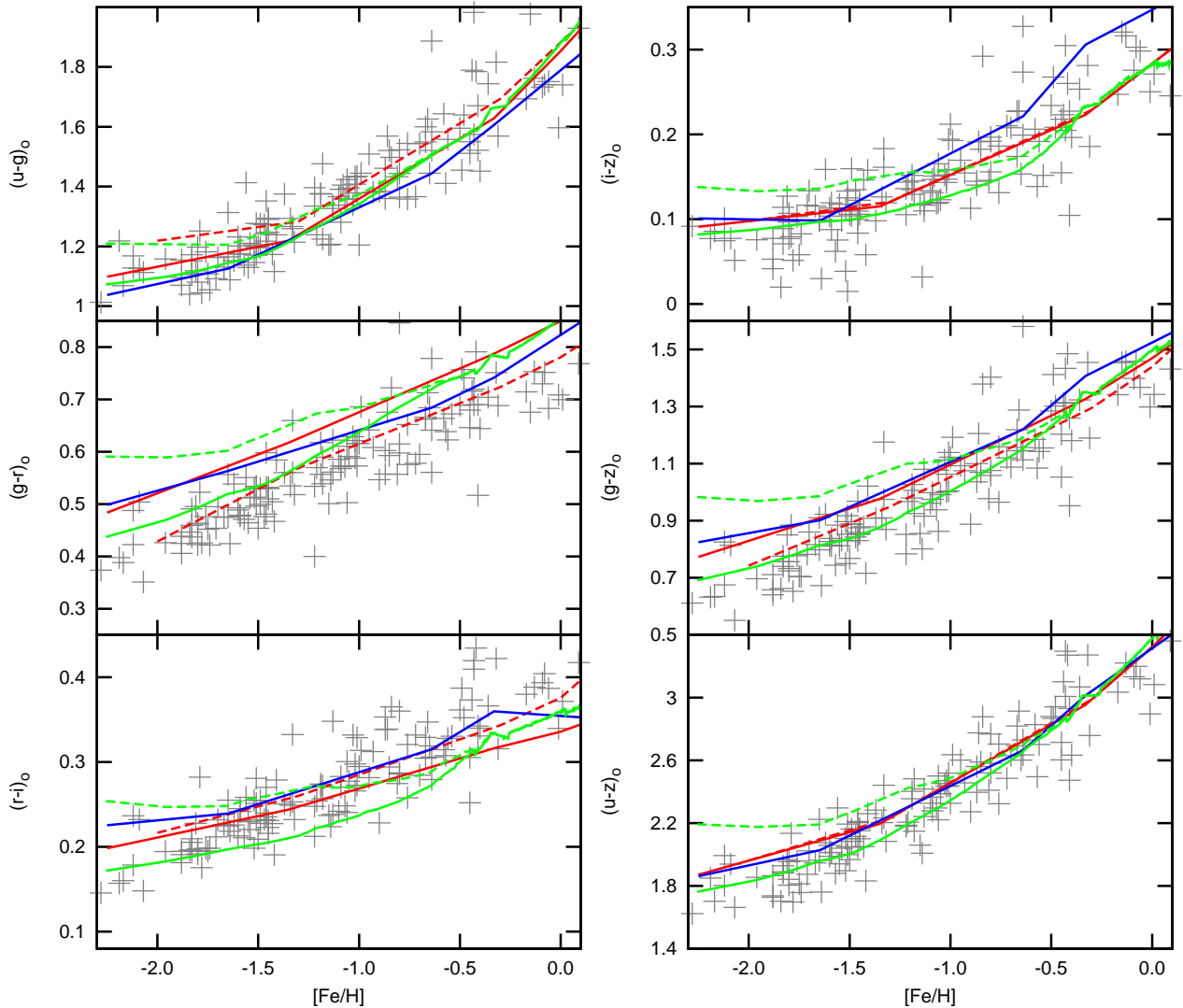} 
 \caption{Metallicity-color relationships of M31's globular clusters. Only low extinction clusters, with E(B-V)$<$0.2, are plotted (gray pluses). Assuming a \citet{Cardelli89} extinction curve, this corresponds to E(\textit{u-g})$<$0.27, E(\textit{g-r})$<$0.21, E(\textit{r-i})$<$0.13, E(\textit{i-z})$<$0.12, E(\textit{g-z})$<$0.46, E(\textit{u-z})$<$0.73. Lines show the predicted colors as a function of metallicity for 12 Gyr SSP models from BC03 (blue), M05 (red), MS11-MILES+M05 (red dashed), Padova (green dashed) and Padova-corrected (green).}
 \label{fig:ugriz_Z} 
\end{figure*}

Figure \ref{fig:ugriz_Z} shows the \textit{ugriz} colors predicted by the SSP models, discussed in section 3, and the intrinsic \textit{ugriz} colors of M31's clusters, as a function of metallicity. The gray pluses show the dereddened colors of M31's clusters with E(B-V)$<$0.2. Higher extinction clusters show greater scatter in their color-metallicity relationships. However, no significant offset is observed between the intrinsic colors of clusters with high and low extinction. The model tracks plotted in this figure are for 12 Gyr SSPs from BC03 (blue line), M05 with a blue-HB morphology (red line), Padova (green dashed line) and corrected Padova (green solid line). A model based on combining the M05 and MS11-MILES model is also included in this figure (dashed red line). However, discussion of this model is saved for section 5.3. The BC03 track is based on a Chabrier IMF while the other models are based on a Kroupa IMF.

The most striking feature, in figure \ref{fig:ugriz_Z}, is that all of the SSP models predict significantly redder colors than observed in \textit{g-r} (with $\Delta$(\textit{g-r})$\sim$0.1). While all models are offset in this color, the corrected Padova model best represents the \textit{g-r}-metallicity relationship, with the other models diverging from the data at low metallicity. However, this model is also offset from the data. The offset in \textit{g-r} is at similar wavelengths to that previously observed between the BC03 and M05 models and the B-V colors of globular clusters \citep[e.g.][]{Maraston05}. However, the effect appears to be more pronounced in the \textit{g-r} filters. This shift in \textit{g-r} is also similar to that observed between these models and SDSS observations of LRGs (see e.g. figure 1 of \citet{Maraston09} and figure 11 of \citet{Conroy10}). Potentially this offset could also be produced by an offset in the metallicity measurements of the clusters. However, we note that systematically reducing the metallicity of the clusters to explain this offset tends to produce new offsets in the other color-metallicity correlations. Possible explanations for the offset in \textit{g-r} are discussed further in the next section.

Better agreement is found between the other colors observed and those predicted by the SSP models. It can be seen that the corrected Padova model represents all of the observed colors better than the original model and is in reasonable agreement with all colors except \textit{g-r} and (to a lesser extent) \textit{r-i}. The BC03 and M05 models predict redder \textit{g-z} colors than observed. These models also predict trends between \textit{r-i} and metallicity that are slightly shallower than that observed. However, it should be noted that the \textit{r-i} color covers a relatively small range, so that the effect, which is small in absolute color offsets, is nonetheless obvious in the plot. The \textit{u-g} colors are generally consistent with the model predictions. We note that the observed relationship between \textit{u-g} and metallicity appears to flatten at low metallicities. However, the other colors are consistent with a linear relationship between the color and metallicity. In particular the \textit{u-z} color shows a strong correlation with metallicity and is consistent with the colors predicted from the BC03, M05 and corrected Padova models.

The red dashed lines in figure \ref{fig:ugriz_Z} represent the combination of the MS11-MILES model (in the \textit{g} and \textit{r}-bands) and the M05 model (in the \textit{u},\textit{i} and \textit{z}-bands). As discussed in section 5.3, this update to the M05 model is in better agreement with the observed \textit{g-r} and \textit{g-z} colors.

\section{Model-data discrepancies}
\label{sec:gr}

Comparisons between the observed and predicted colors of clusters provide a useful diagnostic with which to compare different models and highlight potential improvements. Here we consider some of the possible explanations/remedies for the SSP models predicting redder colors than observed around the \textit{g} and \textit{r}-bands. 

\subsection{Reliability of cluster colors}

\begin{figure*}
 \centering
 \includegraphics[height=170mm,angle=270]{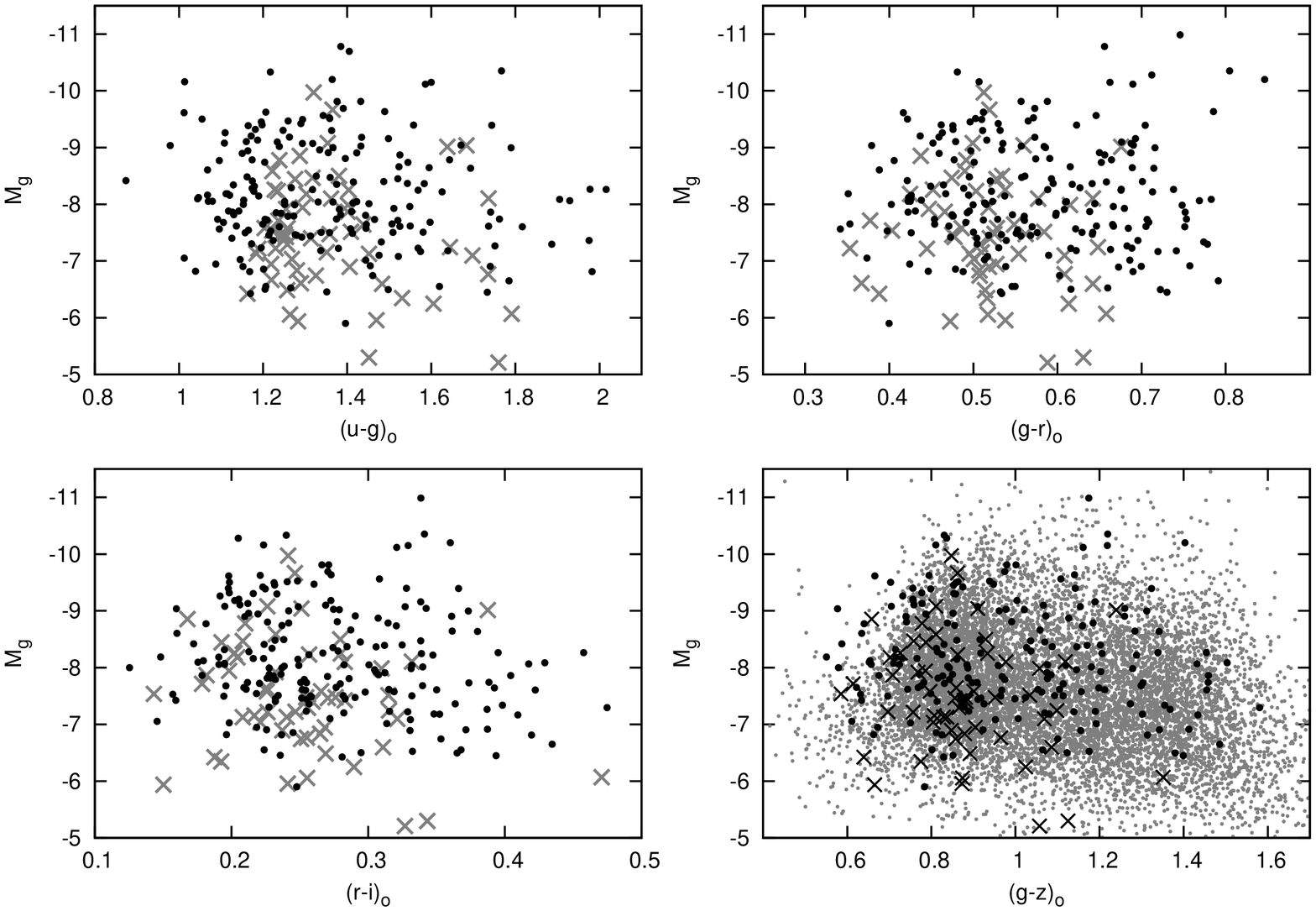} 
 \caption{Colors of clusters in M31 (black points), transformed colors of Milky Way clusters with E(B-V)$<$0.4 (gray/black crosses) and \textit{g-z} color of clusters in the VCS (gray points). Absolute \textit{g}-band magnitudes assume a distance modulus of 24.4 and 31.1 for M31 and the Virgo cluster, respectively.}
 \label{fig:ugriz_comp} 
\end{figure*}

Firstly, we consider the reliability of M31's cluster colors. \citet{Peacock10} have previously compared these data with independent photometry of M31's clusters obtained in the Johnson-Cousins filters. This was done using the photometric transformations presented in \citet{Jester05}. Using these transformations, it was shown that the \textit{ugriz} colors are consistent with the previous UBVR$_{c}$I$_{c}$ photometry of \citet{Barmby00}. A slight offset was found in the \textit{g-r} color (in the sense that \textit{g-r}=\textit{g-r}$_{\rm{Barmby}}$+0.03). However, this offset is within the quoted uncertainty of the transformations used to predict the \textit{g-r} color from the Barmby data. Therefore the two datasets are consistent, to within the accuracy of the transformations. We also note that the slight shift in \textit{g-r} is smaller than, and in the opposite direction to, that observed between M31's cluster colors and the colors predicted from SSP models.

It is also desirable to compare the colors of M31's clusters with the colors of the Galactic globular clusters. Unfortunately, integrated photometry of the Galactic clusters is not currently available in the \textit{ugriz} bands. Instead, the \textit{ugriz} colors of these clusters can be estimated by transforming photometry in the UBVR$_{c}$I$_{c}$ bands \citep[taken from the `December 2010' version of the Harris catalog;][]{Harris96} to the \textit{ugriz} bands using the transformations of \citet{Rodgers06}. These transformations are based on comparison between standard stars that are common to the UBVR$_{c}$I$_{c}$ and \textit{u'g'r'i'z'} filter systems. A comparison between the colors of the Milky Way's and M31's cluster systems is shown in figure \ref{fig:ugriz_comp}. In all colors, fewer red (metal rich) clusters are observed in the Milky Way sample. This is likely due to only including Milky Way clusters with E(B-V)$<$0.4. While this cut ensures the cluster colors are relatively reliable, it also excludes the majority of the metal rich (redder) clusters in the Galaxy. The \textit{g-r} and \textit{r-i} colors are found to be similar between the two cluster systems. It can be seen that M31's clusters are found to extend to bluer \textit{u-g} colors than the Milky Way's. However, the transformation between the \textit{ug} and UB filters is relatively complicated because these filters cover the Balmer break region of the stellar spectra. This produces a jump in the transformation between the \textit{u-g} and U-B colors \citep[see figure 3 of][]{Rodgers06} and is the likely reason for the transformed Milky Way cluster \textit{u-g} colors grouping in a relatively narrow range. It should be noted that the transformations of \citet{Rodgers06} are based primarily on main sequence stars. The integrated emission from these globular clusters will be slightly different to this, including a significant component from cooler red giant stars. It is therefore likely that the difference in \textit{u-g} is due to the photometric transformations used, rather than suggesting an intrinsic difference between the colors of the two cluster systems. 

Relatively little data for other cluster systems has been published in the \textit{ugriz} filters. However, the \textit{g-z} color of a large sample of clusters is available from the HST/ACS Virgo Cluster Survey \citep[VCS; as published by][and obtained from the VizieR archive]{Jordan09}. The VCS obtained photometry through the F475W (SDSS \textit{g}) and F850LP (SDSS \textit{z}) filters of globular clusters in 100 different Virgo cluster galaxies. In the bottom right panel of figure \ref{fig:ugriz_comp} we compare the colors of clusters in the VCS (small gray points) with M31's clusters (black points). It can be seen that M31's clusters span a similar range of \textit{g-z} colors. More red (metal rich) clusters are observed in the VCS dataset, compared with M31. However, this is unsurprising given the large number of clusters in elliptical galaxies in the VCS dataset (these galaxies are known to host relatively large populations of metal rich clusters).

A potential source of uncertainty on the intrinsic colors of M31's clusters is reddening, which varies significantly due to M31 itself. These data were dereddened using the values determined by \citet{Fan08}. However, during the course of this work, a new spectroscopic survey of M31's clusters was published by \citet{Caldwell11}. This provides independent measurements for the metallicities and reddenings of these clusters. Comparing these data with that of \citet{Fan08}, we note only small offsets between the two data sets of $\Delta$[Fe/H]$\sim$0.1 and $\Delta$E(B-V)$\sim$0.05. Adopting the metallicity and reddening values presented by \citet{Caldwell11} actually tends to increase the observed offset between the model and data colors. It is therefore unlikely that the offset (seen in figure \ref{fig:ugriz_Z}) is due to systematic errors in either the reddening correction or spectroscopic metallicities. In this work, we have kept the reddening values presented in \citet{Fan08} because they are found to produce less scatter in both color-metallicity and color-color plots.

Considering these comparisons, it appears unlikely that the shifts we observe between the data and models are due to the calibration of the cluster colors. 

\subsection{Additional stellar populations} 

A potential explanation for offsets between the observed and predicted colors of these clusters is a difference in the underlying stellar populations. Certain phases of stellar evolution are not included in these SPS models, or are known to be relatively poorly treated. One such population are the HB stars, which are notoriously difficult to model. In figure \ref{fig:libraries_gr_Z}, we show the M05 models with red-HB star (dashed red lines) and blue-HB star (solid red lines) morphologies. It can be seen that the addition of blue-HB stars does produce bluer integrated colors and therefore represent the data better, particularly at low metallicity \citep[where clusters are known to have bluer-HB star populations,][]{Sandage60}. However, even for the metal poor clusters, the shift is far less than that required to explain the observed colors. Including more blue/extreme-HB stars could potentially push the models to even bluer colors, but it is unlikely this could produce the (large) observed offset. Blue straggler (BS) stars are also known to be present in globular clusters and are generally excluded from SSP models. These bright, blue, stars would tend to shift the models towards bluer colors. However, clusters typically contain only a small fraction of blue stragglers, so it is unlikely they will produce the large offset seen in most of M31's clusters. For example, if we assume that a cluster contains approximately 1 blue straggler star per 2000M$_{\sun}$ \citep{Knigge09} and that a typical blue straggler has \textit{g}$\sim$2.4 and \textit{g-r}$\sim$0 \citep{Leigh07}, we would expect the addition of such a population to make the model of an old, metal poor cluster bluer by only \textit{g-r}$\sim$0.01. Therefore, while the addition of these stellar populations can produce bluer cluster colors, it appears unlikely that they can produce the large offset observed in M31's clusters. The addition of these hot populations would also be expected to influence the \textit{u-g} color of these clusters. However, relatively good agreement is found between the observed and expected \textit{u-g} colors, compared with the \textit{g-r} colors (see figure \ref{fig:ugriz_Z}).

It has also been proposed that some of the observed offsets between SPS models and LRG observations, can be explained by adding minority populations of metal poor stars \citep{Maraston09,Conroy10} or young stars \citep{Conroy10} to the expected old, metal rich, stellar population. While such populations may be possible in these galaxies, age and metallicity spreads on this scale are not observed in globular clusters. It is therefore very unlikely that the observed colors of M31's clusters are due to minority populations of younger or lower metallicity stars.

\subsection{Stellar libraries}
\label{sec:gr-libraries}

\begin{figure}
 \centering
 \includegraphics[height=86mm,angle=270]{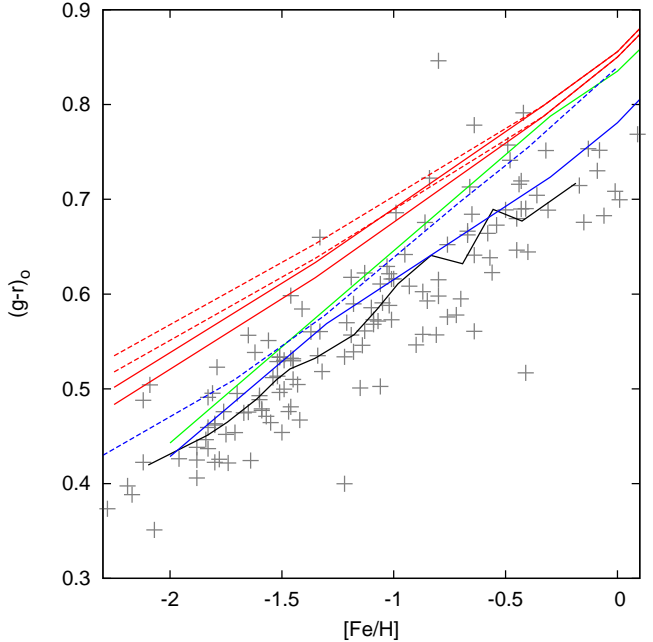} 
 \caption{Comparison between models using different stellar libraries. As in figure \ref{fig:ugriz_Z}, M31's clusters are shown in gray. The black line represents the average color of M31's clusters (based on grouping clusters in to 15 metallicity bins). The red lines show the original models of M05 for 12 Gyr SSPs with blue and red-HB morphologies (solid and dashed lines, respectively). The upper dashed and upper solid red lines are models based on a Salpeter IMF, while the lower red lines are based on a Kroupa IMF. The green line shows the MS11-ELODIE model for an 11 Gyr SSP. The solid blue line shows the MS11-MILES model for a 12 Gyr SSP and the dashed blue line the \citet{Vazdekis10} model.}
 \label{fig:libraries_gr_Z} 
\end{figure}

Another important factor in comparing SPS models with observations is the stellar libraries that are used to convert the models in to observable spectra. Previously, \citet{Maraston09} have proposed that the observed offset between the M05 models and observations of LRGs in the SDSS, could be explained by using updated models based on empirical, rather than the theoretical, stellar libraries. By using these empirical libraries, and including a minority population of low metallicity stars, they found good agreement with the \textit{gri}-band observations of these galaxies at all redshifts considered. However, in a subsequent paper \citet{Conroy10} were unable to explain this offset by using the same empirical libraries in their models. By using these M31 data, we can compare models based on different stellar libraries with the relatively simple stellar populations of globular clusters, over a greater spread in metallicity.

In figure \ref{fig:libraries_gr_Z} we compare the \textit{g-r} colors of M31's clusters with the original SSP models of M05, based on the theoretical Kurucz stellar library (red lines). Considering the different versions of the M05 models, it can be seen that the models which include a blue-HB morphology (solid red lines) provide a better fit to the observations. The M05 models based on a Kroupa IMF (lower red dashed/ solid lines) produce bluer colors than those based on a Salpeter IMF and are therefore in better agreement with our observations. However, all of the M05 models produce redder colors than observed, for all metallicities.

The solid blue line in figure \ref{fig:libraries_gr_Z} shows the colors predicted from the MS11-MILES model. This model is based on the same evolutionary code as the M05 models but uses the empirical MILES stellar library. It can be seen that this model is in relatively good agreement with the observed \textit{g-r} colors for all metallicities. This is consistent with the findings of \citet{Maraston09}, who found that these models also provide a good fit to the SDSS LRG observations. Compared with the theoretical libraries, the MILES stellar library has relatively limited wavelength coverage. Therefore, the MS11-MILES model can not be directly applied to the other SDSS bands studied. However, we note that if these shifts in the \textit{g} and \textit{r}-bands are applied to the original M05 models, it also improves the fit to the observed \textit{g-z} colors of M31's clusters. This updated model is represented in figure \ref{fig:ugriz_Z} as the dashed red line. Reasonable fits are also obtained to the \textit{u-g} color, although the updated models appear to be slightly too red compared with the metal poor clusters. The trend between \textit{r-i} and [Fe/H] remains slightly flat compared with the data. Again we note that \textit{r-i} only covers a relatively small range and that the offset between the data and models is small. However, it is possible that the implied correction to the model spectrum around the \textit{g} and \textit{r}-bands may also have a (smaller) influence on the i and \textit{u}-bands.

An independent stellar population model, that is also based on the MILES stellar library, has been presented by \citet{Vazdekis10}. The \textit{g-r} color of this model was obtained using the data access tool\footnote{provided at http$:$//miles.iac.es/}. The blue-dashed line in figure \ref{fig:libraries_gr_Z} shows the colors predicted from this model. It can be seen that, compared with the M05 models, this model provides a better fit to the metal poor clusters. However, it is still redder than the observed colors, especially at higher metallicities. Given that the \citet{Vazdekis10} and MS11-MILES models are based on the same stellar library, it is interesting that they predict significantly different colors around solar metallicity. The offset between the two models demonstrates the complexity of predicting integrated colors from SPS models, which are highly dependent upon the evolutionary prescriptions used and the implementation of the stellar libraries, as well as the libraries themselves. Our comparison suggests that the use of the MILES library can help to explain the offset between the Maraston models and observations. However, using this library does not appear to explain the similar offset observed between observations and the models of \citet{Vazdekis10} and \citet{Conroy10}. The reason for the observed offset between these models and observations remains unclear. The different predictions of the MS11-MILES and Vazdekis/Conroy models also remains an open issue for the modeling community. However, it may be related to differences in the treatments of the red giant branch/ asymptotic giant branch stars in the Maraston and the (Padova based) Vazdekis/Conroy models.

Figure \ref{fig:libraries_gr_Z} also shows the MS11-ELODIE model. This model is similar to the M05 and MS11-MILES models, but is based on the ELODIE stellar library (green line). It can be seen that this model also provides a better fit to the low metallicity clusters. However, it still predicts redder colors than those observed for the higher metallicity clusters. The difference between this model and the MS11-MILES model must be due to differences between the stellar libraries. A likely reason for this is difficulties in flux calibrating the stellar spectra. Indeed, MS11 demonstrate that significant differences are found between the spectra of stars that are present in both the MILES and ELODIE libraries. The variation in the colors predicted by the otherwise similar M05, MS11-MILES and MS11-ELODIE models highlights the importance of stellar libraries to the predictions of SPS models.

\begin{figure}
 \centering
 \includegraphics[height=86mm,angle=270]{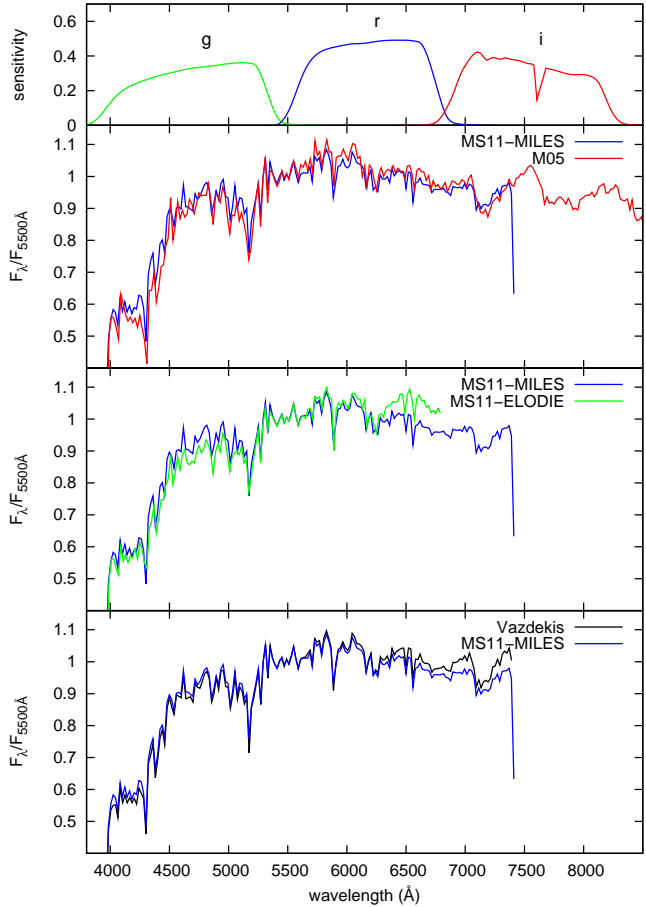} 
 \caption{Spectra of 12 Gyr SSPs with [Fe/H]=+0.00 produced from the models of M05 (red), Vasdekis (black), MS11-MILES (blue) and MS11-ELODIE (green). All models have been scaled by their flux at 5500$\AA$, this is close to the intersection between the \textit{g} and \textit{r}-band filters. The models have also been binned to the spatial resolution of the M05 models. For comparison, the SDSS filter transmission curves are shown in the top panel.}
 \label{fig:ssp_seds} 
\end{figure}

To highlight the differences in these models, we show in figure \ref{fig:ssp_seds} the spectra predicted for stellar populations with solar metallicity. For clarity, we have binned the higher spectral resolution models to that of the M05 models and scaled all fluxes to that at 5500$\AA$ (close to the edge of the \textit{g} and \textit{r}-band filters). This figure demonstrates that the MS11-MILES model produces slightly higher fluxes than the M05 model at most wavelengths covered by the SDSS \textit{g}-band filter. Conversely, this model produces slightly lower fluxes across the SDSS \textit{r}-band filter. This leads to the bluer \textit{g-r} color predicted by this model. In comparison to the MS11-MILES model, the use of the ELODIE library in the MS11-ELODIE model produces slightly less flux in the \textit{g}-band filter. However, the biggest difference between these models is around 6500$\AA$, where the MS11-ELODIE model produces higher fluxes. In the bottom panel of figure \ref{fig:ssp_seds}, it can be seen that, the MS11-MILES and Vazdekis models produce similar fluxes across the \textit{g}-band. However, the Vazdekis model predicts increasingly higher fluxes at longer wavelengths. This produces the the redder \textit{g-r} color predicted by the Vazdekis model, compared with the MS11-MILES model.

\section{Conclusions}

We have demonstrated a significant offset between the colors of M31's globular clusters and those predicted by some SSP models. The biggest discrepancy is found around the SDSS g and \textit{r}-bands. The offset observed in \textit{g-r} is similar to that previously observed in Sloan LRG datasets, but is confirmed here to also effect the relatively simple stellar populations of globular clusters, across a range of metallicities. Identifying this offset in globular clusters suggests that it is unlikely to be due to a small fraction of young stars, as has previously been proposed in the LRGs. Including other minority populations such as extreme/blue-HB stars or blue straggler stars can produce clusters with bluer colors. However, such populations would not be expected to produce the large observed offset in only \textit{g-r}.

We have shown that the \textit{g-r} colors predicted by the MS11-MILES model are consistent the observed cluster colors, for all metallicities. However, the same models based on the theoretical, Kurucz, and empirical, ELODIE, libraries predict significantly redder colors than observed. This suggests that the observed offset in \textit{g-r} can be explained by differences between the stellar libraries used. It should be noted that the Vazdekis model, also based on the MILES library, is still offset in \textit{g-r}. The reasons for the difference between the MS11-MILES and Vazdekis models are currently uncertain and worthy of future investigation. However, we note that the MS11-MILES model is the only model-library combination considered that gives a reasonable representation of the observed colors of M31's clusters.

Considering such shifts between models and observations provides an important tool for refining SPS models. Such offsets should be considered when inferring the properties of globular clusters and galaxies from their integrated emission.

\section*{Acknowledgements}
We would like to thank Dr. Charlie Conroy for refereeing this paper and for providing prompt comments, which were beneficial to the final version. We would also like to thank Claudia Maraston for providing us with an advanced copy of her paper and updated stellar population models. MBP and SEZ would like to acknowledge support from the NASA grant NNX08AJ60G. 

\bibliographystyle{apj_w_etal}
\bibliography{bibliography_etal}

\label{lastpage}

\end{document}